\documentclass[sigconf]{acmart}

\usepackage{amsmath}   
\usepackage{newtxmath} 

\usepackage{graphicx} 
\usepackage{epstopdf}  
\usepackage{xcolor}
\usepackage{colortbl}

\usepackage{array}
\usepackage{multirow}
\usepackage{makecell}
\usepackage{algorithm}
\usepackage{algpseudocode}

\usepackage{tikz}
\usepackage{bm}
\usepackage{seqsplit}
\usepackage{enumitem}

\usepackage{hyperref}

\DeclareMathOperator*{\argmax}{arg\,max}
\DeclareMathOperator*{\argmin}{arg\,min}

\newcommand*\circled[1]{\tikz[baseline=(char.base)]{
            \node[shape=circle,draw,inner sep=0.4pt] (char) {#1};}}

\AtBeginDocument{%
  }

\setcopyright{acmlicensed}
\copyrightyear{2018}
\acmYear{2018}
\acmDOI{XXXXXXX.XXXXXXX}

\acmConference[Conference acronym 'XX]{Make sure to enter the correct
  conference title from your rights confirmation emai}{June 03--05,
  2018}{Woodstock, NY}
\acmISBN{978-1-4503-XXXX-X/18/06}

\begin{document}
\title{PGTuner: An Efficient Framework for Automatic and Transferable Configuration Tuning of Proximity Graphs}

\author{Hao Duan}
\affiliation{%
  \institution{Shanghai Jiao Tong University}
  \city{Shang Hai}
  \country{China}}
\email{bunny\_duanhao@sjtu.edu.cn}

\author{Yitong Song}
\affiliation{%
  \institution{Shanghai Jiao Tong University}
  \city{Shang Hai}
  \country{China}}
\email{yitong\_song@sjtu.edu.cn}

\author{Bin Yao}
\authornote{represents the corresponding author.}
\affiliation{%
  \institution{Shanghai Jiao Tong University}
  \city{Shang Hai}
  \country{China}}
\email{yaobin@cs.sjtu.edu.cn}

\author{Anqi Liang}
\affiliation{%
  \institution{Shanghai Jiao Tong University}
  \city{Shang Hai}
  \country{China}}
\email{lianganqi@sjtu.edu.cn}

\renewcommand{\shortauthors}{Hao Duan, Yitong Song, Bin Yao, \& Anqi Liang}

\begin{abstract}
Approximate Nearest Neighbor Search (ANNS) plays a crucial role in many key areas. Proximity graphs (PGs) are the leading method for ANNS, offering the best balance between query efficiency and accuracy. However, their performance heavily depends on various construction and query parameters, which are difficult to optimize due to their complex inter-dependencies. Given that users often prioritize specific accuracy levels, efficiently identifying the optimal PG configurations to meet these targets is essential. Although some studies have explored automatic configuration tuning for PGs, they are limited by inefficiencies and suboptimal results. These issues stem from the need to construct numerous PGs for searching and re-tuning from scratch whenever the dataset changes, as well as the failure to capture the complex dependencies between configurations, query performance, and tuning objectives.
\par
To address these challenges, we propose PGTuner, an efficient framework for automatic PG configuration tuning leveraging pre-training knowledge and model transfer techniques. PGTuner improves efficiency through a pre-trained query performance prediction (QPP) model, eliminating the need to build multiple PGs. It also features a deep reinforcement learning-based parameter configuration recommendation (PCR) model to recommend optimal configurations for specific datasets and accuracy targets. Additionally, PGTuner incorporates out-of-distribution detection and deep active learning for efficient tuning in dynamic scenarios and transferring to new datasets. Extensive experiments demonstrate that PGTuner can stably achieve the top-level tuning effect across different datasets while significantly improving tuning efficiency by up to 14.69$\times$, with a 14.64$\times$ boost in dynamic scenarios. The code and data for PGTuner are available online at https://github.com/hao-duan/PGTuner.
\end{abstract}


\received{20 February 2007}
\received[revised]{12 March 2009}
\received[accepted]{5 June 2009}

\maketitle

\section{INTRODUCTION}
In recent years, Approximate Nearest Neighbor Search (ANNS) on high-dimensional vectors has gained widespread attention due to its crucial role in many fields, including information retrieval \cite{liu2007survey}, data mining \cite{valkanas2017mining}, recommendation systems \cite{suchal2010full}, and retrieval-augmented generation (RAG) \cite{lewis2020retrieval}. ANNS methods are typically categorized into tree-based \cite{liu2004investigation, wang2013trinary}, hash-based \cite{zheng2016lazylsh, liu2021ei}, quantization-based \cite{wu2017multiscale, matsui2018survey}, and proximity graph (PG) based approaches \cite{HNSW, NSG}. Recent works \cite{wang2021comprehensive, peng2023efficient, zhao2023towards} have shown that PG-based methods deliver the best performance, making them the preferred choice in industrial vector databases like Milvus\footnote{Milvus: \url{https://milvus.io/}} and Weaviate\footnote{Weaviate: \url{https://github.com/weaviate/weaviate}}.
\par
PG-based methods organize vectors as nodes in a graph, connecting each node to its approximate nearest neighbors \cite{wang2021comprehensive}. During index construction, an ANNS algorithm identifies $efC$ candidate neighbors for each node, from which $M$ neighbors are selected for connection using a heuristic pruning algorithm. For query processing, the ANNS algorithm retrieves nodes, stores them in a candidate set of size $efS$, and then returns the top-$k$ nearest nodes from the set. The query performance is evaluated in terms of accuracy (i.e., query recall) and efficiency (i.e., queries per second, QPS), with higher QPS often resulting in lower recall. The trade-off can be adjusted through both construction (e.g., $efC$, $M$) and query parameters (e.g., $efS$). Generally, higher values for these parameters enable the exploration of more high-quality nodes during query processing, improving recall but reducing QPS. Different parameter configurations may achieve similar recall with varying QPS. In practice, users often have specific recall requirements, making the optimization of PG parameters to meet users' recall preferences while maximizing QPS a critical focus \cite{lee2022anna, jiang2023co, aumuller2023overview, yang2024vdtuner}.
\par
Due to the labor-intensive nature of manual tuning and the vast space of parameter combinations, existing research primarily focuses on automating PG configuration tuning. These studies fall into two categories: search-based methods \cite{wang2021comprehensive} and learning-based methods \cite{oyamada2020towards, yang2024vdtuner}. The most commonly used search-based method is GridSearch \cite{wang2021comprehensive}, which defines a large set of candidate configurations and tests them to select the best one. However, this approach is time-consuming due to the need to construct and test numerous PGs, especially when dealing with tens of millions of vectors. VDTuner \cite{yang2024vdtuner} and GMM \cite{oyamada2020towards} are two representatives of learning-based methods. VDTuner improves tuning efficiency by using constrained Bayesian optimization to recommend promising configurations for target recalls. However, it remains inefficient, as it requires constructing PGs to evaluate the recommended configurations during tuning and re-tuning from scratch whenever the dataset changes. In contrast, GMM \cite{oyamada2020towards} uses a pre-trained meta-model to directly predict the query performance of parameter configurations without constructing PGs, and then selects the best configuration for recommendation. Although GMM enables efficient PG tuning on a given dataset, it often recommends unsatisfactory configurations due to its inability to effectively transfer the pre-trained meta-model to the given dataset. Additionally, both VDTuner and GMM fail to effectively model the hidden dependencies between configurations, query performance, and tuning objectives, resulting in unstable and suboptimal configurations across different target recalls. 
\par 
Overall, there are three challenges in automatically tuning PG:
(1) how to effectively capture the hidden dependencies between configurations, query performance, and tuning objectives to reliably recommend high-performance configurations for various target recalls;
(2) how to accurately evaluate PG configurations without actually constructing and testing PGs;
(3) how to effectively and efficiently tune across various and dynamic datasets. 
\par
To address these challenges simultaneously, we propose PGTuner, an efficient framework for automatic and transferable configuration tuning of PGs. For the first challenge, PGTuner trains a parameter configuration recommendation (PCR) model using deep reinforcement learning (DRL) \cite{wang2022deep} to effectively capture the complex dependencies described above, enabling it to consistently recommend high-performance configurations for different target recalls. For the second and third challenges, we design a query performance prediction (QPP) model to avoid constructing PGs and leverage model transfer techniques for generalization to different datasets. Specifically, the PCR model considers generalizable tuning strategies, allowing it to build on small-scale, low-dimensional base datasets and be rapidly transferred to more complex target datasets through fine-tuning. The QPP model enables effective transfer by iteratively retraining with data collected from the target dataset, using deep active learning (DAL) \cite{li2024survey}. To further enhance transfer efficiency, PGTuner also uses Out-of-Distribution (OOD) detection to assess the similarity between the given dataset and base datasets during each iteration. If similarity is detected, the transfer process can be terminated early, significantly reducing tuning time.
\par
In summary, we make the following contributions: 
\begin{itemize}[leftmargin=*]
    \item We propose PGTuner, a transferable auto-tuning framework for PG, which can achieve effective and efficient tuning on any given dataset and target recall.
    \item A DRL-based PCR model is introduced to capture the complex inter-dependencies between configurations, query performance, and tuning objectives. It learns a generalizable and high-quality tuning strategy and can stably recommend high-performance configurations across different datasets and target recalls.
    \item A QPP model for quickly evaluating PG configurations is proposed, together with an OOD detection- and DAL-based model transfer method to ensure effective and efficient transfer of the QPP model to different datasets.
    \item Extensive experiments on real-world datasets demonstrate that PGTuner can stably achieve the top-level tuning effect across different datasets while significantly improving tuning efficiency. Compared with the state-of-the-art baseline, PGTuner increases tuning efficiency by up to 14.69$\times$, and also achieves a 14.64$\times$ boost in the dynamic scenario where the dataset size scales up.
\end{itemize}

\begin{figure}[t]
  \centering
  \includegraphics[width=\linewidth]{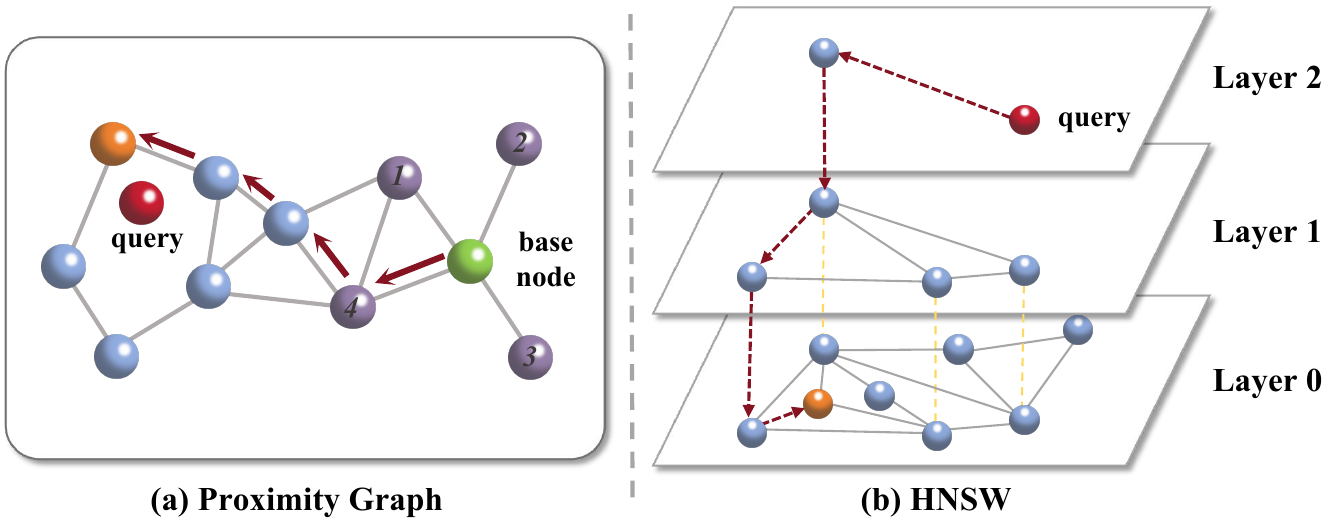}  
  \caption{Examples of proximity graph and HNSW index.}
  \label{fig:PG_graph}
\end{figure}

\section{PRELIMINARIES}
In this section, we introduce the proximity graph and define the problem of PG configuration tuning.
\subsection{Proximity Graph}
A proximity graph (PG) is a data structure used for approximate nearest neighbor search. It treats vectors as graph nodes, with edges connecting neighboring nodes based on their distances evaluated by a distance function (e.g., Euclidean distance). As shown in Figure \ref{fig:PG_graph} (a), the four nodes (numbered 1–4) connected to the green node are its neighbors. During PG construction, for each inserted node $i$, the search starts at an initial base node (e.g., the green node) and examines the distances from $i$ to the current base node and its neighbors. The closest node is then selected as the next base node, while dynamically recording $efC$ candidate approximate nearest neighbors (NNs) for $i$. The search is terminated when no new closer node to $i$ is found. A heuristic pruning algorithm is then applied to select M candidate nodes as the final neighbors of $i$. During query processing, for a query vector $q$, the search begins from an initial node and proceeds in the same way while continuously maintaining $efS$ candidate approximate NNs for $q$. The search stops when no closer node is identified, and the top-$k$ NNs are returned.
\par
\begin{figure}[t]
  \centering
  \includegraphics[width=\linewidth]{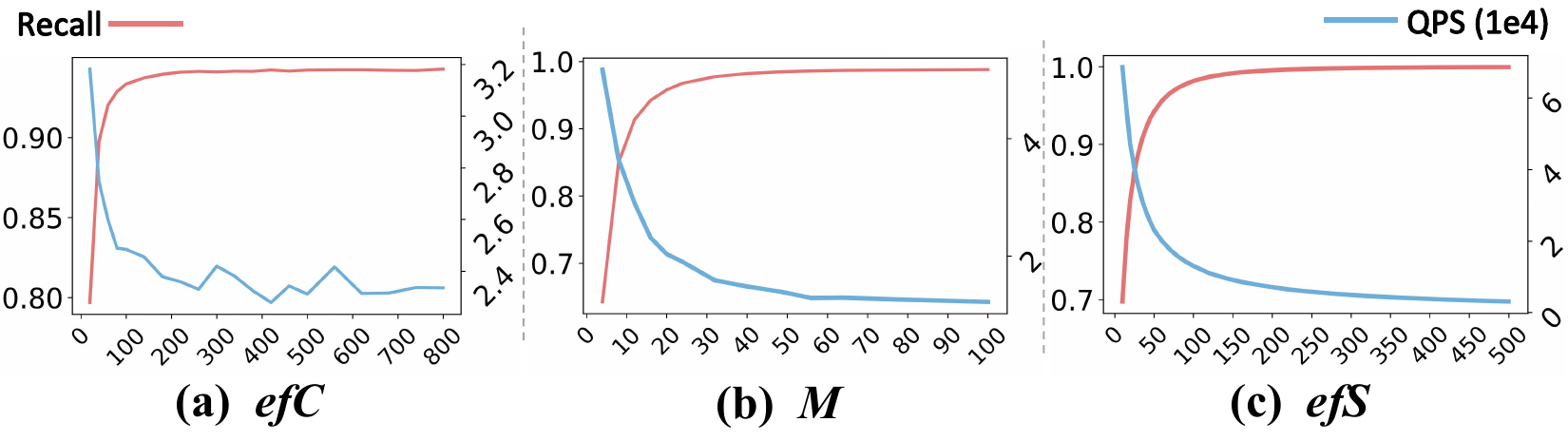}  
  \caption{(a), (b) and (c) show the variations of recall and QPS with $efC$, $M$ and $efS$ on the SIFT1M dataset, respectively. The selected configuration is (500, 16, 50), fixing two of them and varying the third one each time.}
  \label{fig:pc}
\end{figure}

The query performance of PG is highly sensitive to its construction and search parameters. In general, increasing $efC$ improves the neighbor precision of each node, which helps identify more true NNs during query processing, thereby enhancing recall. However, this improvement may increase the number of nodes explored, leading to decreasing QPS. In contrast, increasing $M$ expands each node’s neighbors, which enhances recall because more potential true NNs might be explored, but decreases QPS. Similarly, Larger $efS$ allows for exploring more nodes, boosting recall at the cost of decreased QPS. Figure \ref{fig:pc} shows that these parameters significantly influence the query performance of PG and there is a trade-off between the recall and QPS.
\par
In this paper, we introduce our tuning framework PGTuner based on HNSW \cite{HNSW}, a widely used PG index. HNSW is structured as a hierarchical proximity graph, as shown in Figure \ref{fig:PG_graph} (b). Each layer stores a proximity graph with nodes that are a subset of those in the layer below, and the bottom graph contains all the vectors. During the HNSW construction, the highest level $l$ for each inserted node $i$ is determined, and $i$ is inserted into graphs from level $l$ down to the bottom. The insertion process in each level is identical to that described above. For query processing, HNSW starts from a random node on the top level and recursively traverses downward to find a node close to the query vector $q$ in the bottom graph. It then continues to search in the bottom graph from this node and returns the $k$NNs of $q$. The query performance of HNSW is also determined by $efC$, $M$ and $efS$. Note that, although the PGTuner is introduced based on HNSW, it can be easily extended to other PGs.

\subsection{PG Configuration Tuning}
Consider a PG with $n$ tunable parameters, \(p_1, p_2, \cdots, p_n\), where the first $n_C$ parameters are construction parameters and the rest are search parameters. A parameter configuration $\theta$ is defined as (\(p_1, p_2, \cdots, p_n\)), where (\(p_1, p_2, \cdots, p_{n_C}\)) represents the construction parameter configuration $\theta_C$, while (\(p_{n_{C+1}}, \cdots, p_n\)) is the search parameter configuration $\theta_S$. Assuming the value space of the $i$-th parameter is $\Theta_i$, then the entire configuration space is \(\Theta = \Theta_1 \times \Theta_2 \times \cdots \times \Theta_n\). Given a dataset containing a base vectors dataset (BD) and a query vectors dataset (QD), and a parameter configuration $\theta$, we first construct a PG on BD based on $\theta_C$, then execute all queries in QD using $\theta_S$ to obtain the corresponding query performance. As users often have specific recall requirements, it is essential to find the optimal PG configurations that achieve the desired recalls while maximizing QPS. In this work, we define the PG configuration tuning problem as follows.
\par
\noindent \textbf{Definition 1} \textbf{\textit{(PG Configuration Tuning, PGCT)}.} \textit{Given a dataset} \( X \in \mathbb{R}^d \) \textit{, a PG with a configuration space $\Theta$, and a target recall} $rec_{t}$ \textit{, PGCT aims at finding the optimal configuration} \( \theta^{\ast} \in \Theta \) \textit{such that the corresponding recall} \({rec}_{\theta^{\ast}}\) \textit{reaches} \( rec_{t}\) \textit{and the queries per second} \({QPS}_{\theta^{\ast}}\) \textit{achieves the maximum. This can be formulated as follows.} 
\begin{equation}
\begin{aligned}
 \theta^\ast &= \argmax_{\theta \in \Theta} ({QPS}_{\theta}) \\
\text{s.t.} & \quad {rec}_{\theta} \geq rec_{t}
\end{aligned}
\end{equation}

\begin{figure*}[tbp]
    \centering
    \includegraphics[width=\textwidth]{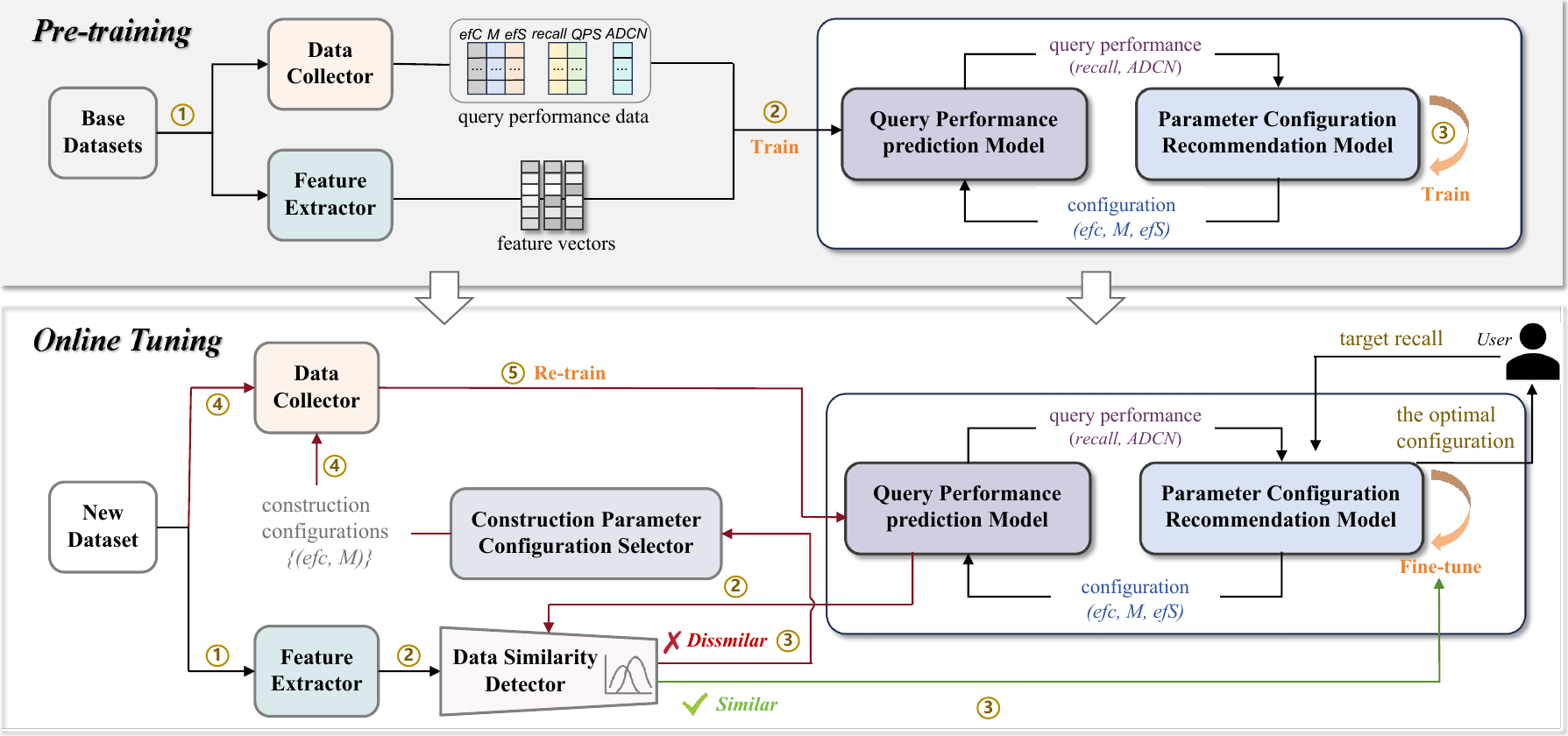} 
    \caption{Architecture and Working Mechanism of PGTuner.}
    \label{fig:wm}
\end{figure*}

\section{OVERVIEW OF PGTUNER}
In this section, we propose our automatic PG configuration tuning framework PGTuner. The compositions and working mechanism of PGTuner are presented in Figure \ref{fig:wm}. 

\subsection{Composition of PGTuner}
PGTuner consists of six core components: (1) Data Collector, which gathers query performance data on a given dataset; (2) Feature Extractor, which extracts the dataset's features; (3) Query Performance Prediction (QPP) model, which predicts the query performance of configurations; (4) Parameter Configuration Recommendation (PCR) model, which recommends the optimal configuration for the dataset and a target recall; (5) Data Similarity Detector, which detects the similarity between datasets; and (6) Construction Parameter Configuration Selector, which adaptively selects a small set of construction parameter configurations to update the QPP model. These components are discussed in Section 4.

\subsection{Working Mechanism}
As shown in Figure \ref{fig:wm}, the working mechanism of PGTuner is divided into two phases: pre-training and online tuning. 

\subsubsection{Pre-training}
The pre-training phase aims to pre-train high-performance QPP and PCR models on a few small-scale and low-dimensional datasets (Referred to as base datasets). PGTuner first uses the Data collector to collect query performance data (referred to as base query performance data) of predetermined configurations on base datasets. Concurrently, PGTuner employs the Feature Extractor to extract representative features from the dataset as its feature vector (\circled{1}). It then uses the base query performance data and feature vectors to train a high-precision QPP model (\circled{2}). Finally, PGTuner utilizes the QPP model to train a powerful PCR model (\circled{3}). The PCR model iteratively recommends configurations for nine default target recalls (ranging from 0.85 to 0.99). The recommended configurations in each iteration are fed into the QPP model to obtain corresponding query performance. The query performance is then fed back to the PCR model to compute rewards, which are used to adjust its tuning strategy and proceed with the next recommendation. Through this training process, the PCR model learns a high-quality and generalizable tuning strategy.

\subsubsection{Online Tuning}\label{online_tuning}
During the online tuning phase, PGTuner adaptively recommends the optimal parameter configuration on a given new dataset and specified target recall. PGTuner first employs the Feature Extractor to extract the feature vector of the dataset (\circled{1}). It then uses the Data Similarity Detector to assess the similarity between the new dataset and base datasets based on their feature vectors and the pre-trained QPP model (\circled{2}). If the result indicates similarity, the pre-trained PCR model is fine-tuned in the same way as during the pre-training phase, where the configuration that reaches the target recall while achieving the maximum QPS during fine-tuning is considered optimal (\circled{3}). If the result indicates dissimilarity, the model transfer process is initiated to adapt the pretrained QPP model to the new dataset by iterative re-training. In each iteration, PGTuner first selects a small number of construction parameter configurations using the Construction Parameter Configuration Selector (\circled{3}). The Data Collector then collects query performance data for these Construction configurations (\circled{4}). The collected data is mixed with all base query performance data to re-train the QPP model (\circled{5}). Finally, the updated QPP model is used for a new round of data similarity detection (\circled{2}). This process is repeated until the detection result is similar or the maximum number of iterations is reached. Ultimately, the pre-trained PCR model is fine-tuned in the same way (\circled{3}). In addition, once the QPP model is transferred to the new dataset, any changes in the target recall can be quickly addressed by fine-tuning the PCR model.

\section{DESIGN AND IMPLEMENTATION OF PGTUNER}
In this section, we describe the design of the six core components of PGTuner in detail.

\subsection{Data Collector}
The Data Collector collects query performance data for training the QPP and PCR models, which includes parameter configurations (e.g., ($efC, M, efS$)) and their corresponding query performance (e.g., recall and QPS). It takes the construction parameter configuration as input and returns the corresponding query performance data. For each given construction configuration, the Data Collector first constructs the corresponding PG, then iterates through all predefined search parameter configurations to repeatedly execute queries and record the corresponding query performance. During the traversal process, the query process is terminated when both the values of search parameters and query recall reach preset thresholds, thereby speeding up the data collection process.
\par
Note that QPS cannot be used directly for model training, as it is related to device performance. Training models with QPS data collected from a single device will impair their transferability to other devices. Considering that the time spent on distance computations dominates query processing \cite{li2020improving} and is proportional to the number of distance calculations (DCN), we use DCN as a surrogate for QPS in model training because it is independent of device performance. To mitigate the impact of dimensional differences across different QD sizes, we further normalize the DCN by dividing it by the QD size to obtain the average DCN (ADCN). A higher QPS is associated with a lower ADCN. Therefore, maximizing QPS in the $PGCT$ problem corresponds to minimizing ADCN.

\subsection{Feature Extractor}\label{sec:feature_extractor}
Feature extractor extracts representative features of a dataset that are relevant to query performance. There are two important types of characteristics that can describe a dataset, which are general measures and hardness measures \cite{oyamada2020towards}. General measures contain basic information about the dataset, such as the dataset cardinality and vector dimensionality \cite{li2019approximate}. The hardness measure is related to the complexity of the dataset. The widely used hardness measure is the local intrinsic dimension (LID) \cite{li2019approximate, HNSW, wang2021comprehensive}. A dataset with a larger LID value implies it is harder than others \cite{li2019approximate}, typically requiring longer search time to achieve the same recall.
\par
In this work, for general measures, we extract the cardinality and vector dimensionality of BD, which are related to the query recall and ADCN. Larger cardinality and dimensionality of BD generally increase its difficulty, leading to a decline in recall \cite{wang2021comprehensive}. Moreover, the cardinality of QD is also likely to influence the query recall. For instance, when the cardinality increases with the addition of more difficult query vectors, the recall decreases. For hardness measures, in addition to LID, we also designed two other features involving BD and QD. First, we observe that the vectors that are closer to their neighbors are more likely to form a well-connected clique. Therefore, when the vector is one of the $k$ nearest neighbors (NNs) of a query vector, it can be accessed from multiple directions through its neighbors, improving the query efficiency. Additionally, if the difference between the distance from a query vector to its $k$NNs in BD and the distance to its non-$k$NNs is large, it is easy to distinguish them accurately during the query, thus enhancing query efficiency and accuracy. Based on these observations, we first calculate the sum of distances of each vector in BD with its $k$NNs, denoted as $DS$. We also calculate the ratio between the average distance from a query vector to its $k$NNs in BD and the average distance to all other non-$k$NNs, denoted as $DR$. We extracted the statistics of $DS$ and $DR$ similar to \cite{oyamada2020towards}, including minimum, mean, maximum, and standard deviation. The extracted dataset features are as follows:
\begin{equation}
\begin{aligned}
 DF = &\left( C_B, C_D, D, LID, {DS}_{min}, {DS}_{mean}, {DS}_{max}, \right. \\
     &\left. {DS}_{std}, {DR}_{min}, {DR}_{mean}, {DR}_{max}, {DR}_{std}\right.)
\end{aligned}
\label{eq:DF}
\end{equation}
\par
\noindent where $C_B$ and $C_D$ denote the cardinality of BD and QD, respectively; $D$ represents the dimensionality; $LID$ refers to the LID measure; \({DS}_{min}, {DS}_{mean}, {DS}_{max}\) are the minimum, mean, maximum, and standard deviation of $DS$, similar to those used in $DR$. The value of $k$ is set to 10 according to the preliminary experimental results.

\subsection{Query Performance Prediction Model}
The QPP model is used to predict the query performance of parameter configurations on different datasets. We build a neural network composed of 4-layer MLPs \cite{rumelhart1986learning} as the QPP model, as MLP has the advantages of strong expressive power and fast calculation speed. The model input is composed of parameter configuration and the dataset's feature vector, and the output is recall and ADCN. The loss function is the mean squared error (MSE) loss. 
\subsection{Parameter Configuration Recommendation Model}
Configuration tuning can be modeled as a sequential decision-making task. Considering that deep reinforcement learning (DRL) has demonstrated strong capabilities in tackling such tasks, with outstanding performance in database configuration tuning \cite{li2019qtune, zhang2019end, cai2022hunter}, we train a DRL-based PCR model to recommend the optimal parameter configurations. During the tuning process, the tuned PG and the QPP model form the environment, while the PCR model acts as the agent in DRL. The PCR model takes the environment state as input and outputs a parameter configuration. The corresponding query performance predicted by the QPP model causes changes in the environment state, which are used to compute rewards. The agent then updates its tuning strategy based on the state and reward, aiming to recommend better configuration. This process iterates until the model stably converges, indicating that the model has learned the optimal tuning strategy. In this process, the state and reward are crucial as they directly affect the learning speed and performance of the PCR model. Therefore, we provide a detailed description of the design of the state and reward function. Additionally, we introduce a post-processing operation to optimize the recommended configurations.
\subsubsection{State}
It is imperative to provide comprehensive and accurate environmental information to enable the agent to make correct decisions. Given that the agent needs to recommend the configuration that achieves the target recall while minimizing ADCN, it needs to know the target recall as well as the recall and ADCN of each recommended configuration. Furthermore, to enable the agent to leverage previous tuning experiences, it should be provided with the current best configuration and corresponding query performance at each iteration. This helps guide the agent toward promising tuning direction and motivates it to continuously recommend better configurations. Additionally, the agent should not know specific dataset features and explicit target recall value, which encourages it to learn a generalizable tuning strategy. Based on these principles, the state is designed as follows:
\begin{equation}
\begin{aligned}
 S = (\theta_l, \theta_b, {rec}_{\theta_l}, {ADCN}_{\theta_l}, \Delta {rec}_{l,t}, \Delta {rec}_{b,t}, \Delta {rec}_{l,b}, \Delta {ADCN}_{l,b})
\end{aligned}
\label{eq:state}
\end{equation}
where $\theta_l$ and $\theta_b$ represent the last and the current best configurations; ${rec}_{\theta_l}$ and ${ADCN}_{\theta_l}$ denote the recall and ADCN of $\theta_l$; $\Delta {rec}_{l,t}$, $\Delta {rec}_{b,t}$, $\Delta {rec}_{l,b}$ and $ \Delta {ADCN}_{l,b}$ are the performance changes, which can be calculated as follows:
\begin{equation}
\begin{aligned}
 &\Delta {rec}_{l,t} = {rec}_{\theta_l} - {rec}_t \\
&\Delta {rec}_{b,t} = {rec}_{\theta_b} - {rec}_t \\
&\Delta {rec}_{l,b} = {rec}_{\theta_l} - {rec}_{\theta_b} \\
&\Delta {ADCN}_{l,b} = \frac{-{ADCN}_{\theta_l} + {ADCN}_{\theta_b}}{{ADCN}_{\theta_b}}
\end{aligned}
\label{eq:performance_change}
\end{equation}
where ${rec}_t$ represents the target recall, while ${rec}_{\theta_b}$ and ${ADCN}_{\theta_b}$ are the recall and ADCN of $\theta_b$. Using $\Delta {rec}_{b,t}$, $\Delta {rec}_{l,b}$ and $\Delta {ADCN}_{l,b}$ instead of ${rec}_{\theta_b}$ and ${ADCN}_{\theta_b}$ in $S$ can help reduce information redundancy and enhancing the agent's learning efficiency and effectiveness.
\par
The aforementioned design helps the agent fully comprehend the performance changes caused by recommended configurations, thereby enabling it to make better recommendations. Note that both the $\theta_l$ and $\theta_b$ are the configurations with the minimal values in the first iteration. During the pre-training phase, the agent is trained with nine preset target recalls (from 0.85 to 0.99). In the online tuning phase, the agent can efficiently recommend optimal configurations for target recalls specified by users.
\subsubsection{Reward Function}
The reward function calculates rewards for each recommended configuration to encourage the agent to optimize its tuning strategy continuously. Due to the tuning objective is to achieve the target recall while minimizing ADCN, we devise the following tuning approach: 
\par
{
\setlength{\parindent}{0pt} 

(1) Initially, the agent should progressively recommend configuration with increasing recall until the recall meets the target recall, which is fundamental to achieving the above objective.

(2) Once the agent has recommended a configuration whose recall reaches the target recall, it should then progressively recommend a configuration with a smaller ADCN in each iteration.

(3) During the tuning process of (2), the agent must ensure that the recall of each recommended configuration should not be lower than the target recall, otherwise the optimization process is meaningless.
}
\par
On the basis of the reward function proposed by \cite{zhang2019end}, we design a new reward presented below to realize the above tuning idea. 
\par
At the current iteration, we first calculate the performance changes caused by the last recommended configuration $\theta_l$ according to Equation \ref{eq:performance_change}, and we set different computational conditions below:
\begin{equation}
\begin{aligned}
cond1 &= \Delta rec_{l,t} < 0 \text{ and } \Delta rec_{b,t} < 0 \\
cond2 &= \Delta rec_{l,t} \geq 0 \text{ and } \Delta rec_{b,t} < 0 \\
cond3 &= \Delta rec_{l,t} < 0 \text{ and } \Delta rec_{b,t} \geq 0 \\
cond4 &= \Delta rec_{l,t} \geq 0 \text{ and } \Delta rec_{b,t} \geq 0 \text{ and } \Delta ADCN_{l,b} \geq 0 \\
cond5 &= \Delta rec_{l,t} \geq 0 \text{ and } \Delta rec_{b,t} \geq 0 \text{ and } \Delta ADCN_{l,b} < 0
\end{aligned}
\label{eq:condition}
\end{equation}
Then the equation for calculating reward $r$ is as follows:
\begin{equation}
r = 
\begin{cases} 
-(1 - \Delta rec_{l,t})^2 +1 & if \quad cond1 \\
(1 + \Delta rec_{l,t})^2 \cdot (1 + \Delta rec_{l,b}) & if \quad cond2 \\
-(1 - \Delta rec_{l,t})^2 \cdot (1 - \Delta rec_{l,b}) & if \quad cond3 \\
(1 + \Delta ADCN_{l,b})^2 - 1 & if \quad cond4 \\
-(1 - \Delta ADCN_{l,b})^2 + 1 & if \quad cond5 \\
\end{cases}
\label{eq:reward}
\end{equation}
As illustrated in Equations \ref{eq:condition} and \ref{eq:reward}, when either \({rec}_{\theta_l}\) or \({rec}_{\theta_b}\) is below \({rec}_t\), the reward is calculated using only these three recalls. This incentivizes the agent to recommend or maintain the configuration that achieves \({rec}_t\). Once both \({rec}_{\theta_l}\) and \({rec}_{\theta_b}\) reach \({rec}_t\), the reward calculation shifts to utilize only \({ADCN}_{\theta_l}\) and \({ADCN}_{\theta_b}\). This adjustment encourages the agent to continuously recommend configurations with smaller ADCN. Through the above tuning process, the agent can recommend the optimal configuration with the smallest ADCN while achieving the target recall.

\subsubsection{Post-processing for the Recommended Configuration}\label{sec:post_processing}
Due to the prediction errors of the QPP model, the query recall of the recommended optimal configuration may either fall below the target or exceed it by a large margin, meaning this configuration does not achieve the optimal performance. To address this issue, we designed a simple but effective post-processing method. For the recommended configuration, we first construct the PG based on the construction configuration, then execute queries with the search configuration to obtain the corresponding query recall. If the recall is below the target, we iteratively increase the values of the search configuration by a fixed step size and replay the queries until the recall reaches the target. At that point, the corresponding configuration is considered the ultimate recommended configuration. Conversely, if the initial recall exceeds the target, we iteratively decrease the values of the search configuration and execute queries. The process is terminated once the recall falls below the target, at which point the search configuration from the last iteration is returned. Then the configuration composed of the construction configuration and the returned search configuration is the ultimate recommended configuration. This process only needs to be executed once to optimize the optimal configuration ultimately recommended by the PCR model, and it can typically be completed within a very short time.
\par
In this work, we utilize the Twin Delayed Deep Deterministic Policy Gradient (TD3) algorithm \cite{fujimoto2018addressing}, which can stably and efficiently learn high-quality tuning strategy. The actor and critic networks of the TD3 are both neural networks composed of 4-layer MLPs.

\subsection{Data Similarity Detector}
The Data Similarity Detector detects the similarity between a new dataset and base datasets. To achieve efficient and accurate detection, we design a detection algorithm based on an Out-of-Distribution detection algorithm proposed in \cite{sun2022out}, which aims at detecting whether test data falls within the distribution of training data. It computes the distance between the test data's embedding and its $k$-th nearest neighbor in the training data's embeddings generated by a model. If this distance is below a predefined threshold, the test data can be considered similar to the training data. 
\par
However, one issue for performing detection is the lack of explicit test data. To address this problem, the Data Similarity Detector combines the new dataset's feature vector with all candidate parameter configurations to generate corresponding input feature vectors as the test data, denoted as \(\bm{F}^{in}_n\). It then feeds these feature vectors into the QPP model to generate output feature vectors from the model's penultimate layer, denoted as \(\bm{F}^{out}_n\), where each feature vector in it is represented as \(\bm{f}^{out}_n\). Similarly, \(\bm{F}^{in}_b\) represents the input feature vectors of the base training data, and the corresponding output feature vectors are generated in the same way, denoted as \(\bm{F}^{out}_b\),  where each feature vector is represented as \(\bm{f}^{out}_b\). Note that the base training data is assumed to be accessible during online tuning, which is consistent with related researches on out-of-distribution detection \cite{yang2024generalized} and active learning \cite{li2024survey}.
\par
Upon obtaining \(\bm{F}^{out}_b\) and \(\bm{F}^{out}_n\), the Data Similarity Detector conducts similarity detection as shown in Algorithm \ref{alg:dad}.
\par
{
\setlength{\parindent}{0pt} 
(1) Calculate the distance between each \(\bm{f}^{out}_b\) and its $k$-th nearest neighbor in \(\bm{F}^{out}_b\), then sort all these distances in ascending order and set the 95th percentile as the distance threshold (Lines 2--3).

(2) Calculate the average distance between all \(\bm{f}^{out}_n\) in \(\bm{F}^{out}_n\) and their $k$-th nearest neighbors in \(\bm{F}^{out}_b\) (Line 5). If this average distance is below the distance threshold, the new dataset can be considered similar to the base datasets; otherwise, it is dissimilar (Line 6).
}
\par

\begin{algorithm}[t]
\caption{Data Similarity Detection} 
\label{alg:dad}
\begin{algorithmic}[1] 
\renewcommand{\algorithmicrequire}{\textbf{Input}} 
\Require input feature vectors of base datasets \(\bm{F}^{in}_b\),  input feature vectors of new dataset \(\bm{F}^{in}_n\), QPP model \(\phi\) 
\State For \(\bm{f}^{in}_{b_i}\) in \(\bm{F}^{in}_b\), generate output feature vectors by \(\phi\)
\Statex \(\bm{F}^{out}_b = (\phi(\bm{f}^{in}_{b_1}), \phi(\bm{f}^{in}_{b_2}), \dots, \phi(\bm{f}^{in}_{b_{N}})) = (\bm{f}^{out}_{b_1}, \bm{f}^{out}_{b_2}, \dots, \bm{f}^{out}_{b_{N}})\)
\State For \(\bm{f}^{out}_{b_i}\) in \(\bm{F}^{out}_b\), calculate the distance between it and its \(k\)-th nearest neighbor 
\Statex \(\bm{Dist} = (d_1, d_2, \dots, d_{N})\)
\State Resort \(\bm{Dist}\) in ascending order and set the 95th percentile as the distance threshold \(d_{tr}\)

\State For \(\bm{f}^{in}_{_i}\) in \(\bm{F}^{in}_n\), obtain output feature vectors in the same way
\Statex \(\bm{F}^{out}_n = (\bm{f}^{out}_{n_1}, \bm{f}^{out}_{n_2}, \dots, \bm{f}^{out}_{n_M})\)
\State Calculate the average distance \(\bar{d}\) between all \(\bm{f}^{out}_n\) in \(\bm{F}^{out}_n\) and their \(k\)-th nearest neighbors in \(\bm{F}^{out}_b\)
\State Check if \(\bar{d} \leq d_{tr}\), return True or False
\renewcommand{\algorithmicrequire}{\textbf{Output}} 
\Require The result of the data similarity detection (True/False)
\end{algorithmic}
\end{algorithm}

According to the preliminary experimental results, increasing $k$ causes the threshold $d_{tr}$ to grow more rapidly. An excessively large threshold causes incorrect detections, hindering the effective transfer of the QPP model to the new dataset and further preventing the PCR model from making high-quality recommendations. Therefore, $k$ is set to 1 for the most stringent detection.

\subsection{Construction Parameter Configuration Selector}
During the online tuning phase, if a new dataset is detected to be dissimilar to the base datasets, it is essential to collect some query performance data on the new dataset to update the pre-trained QPP model, enabling it to predict accurately on both the base and new datasets. This process involves two key challenges: (1) which parameter configurations to select for data collection, and (2) how many configurations are needed. If the selected configurations are not representative, the collected data may not effectively facilitate transferring the pre-trained QPP model to the new dataset. On the other hand, selecting too few configurations results in insufficient data, which may also cause the updated model to fail to predict accurately on the new dataset. Conversely, choosing too many configurations will significantly increase the time required for data collection, thus extending the tuning time. 

Considering that obtaining the query performance for given configurations can be viewed as a data labeling task, the DAL technique is well-suited to address this task. DAL aims to achieve strong model performance with fewer labeled samples, which is suitable for adaptively selecting representative construction configurations to enable effective model transfer. Therefore, we borrow from the core-set selection algorithm \cite{sener2017active} to solve the above challenges, which iteratively selects data (i.e., core-set) from the unlabeled data with the largest nearest neighbor distance (NND) to the current labeled data. This can effectively ensure that the model updated with the core-set can perform well on the remaining unlabeled data. 

\begin{algorithm}[t]
\caption{Construction Parameter Configurations Selection}
\label{alg:cpcs}
\begin{algorithmic}[1]
\renewcommand{\algorithmiccomment}[1]{/* #1 */}
\renewcommand{\algorithmicrequire}{\textbf{Input}}
\Require \(N_C\) candidate construction parameter configurations \(\bm{\theta^C}\),
\Statex \hspace{3mm} base datasets' features \(\{DF_b\}\), new dataset's feature \(DF_n\); 
\Statex \hspace{3mm} initial input feature vectors: \(\bm{F}^{in}_b\), \(\bm{F}^{in}_n\); 
\Statex \hspace{3mm} QPP model \(\phi\), base training data \(\mathcal{T}\), selection rounds R

\For{$i \in \{1, \cdots ,\text{R}\}$} 
    \State Initialize the selected construction parameter configura-
    \Statex \hspace{4mm} tions U = \(\varnothing\), corresponding input feature vectors are \(\bm{F}^{in}_u\)
    \State Generate output feature vectors \(\textbf{F}^{out}_b\) corresponding to \(\textbf{F}^{in}_b\)
    \For{$j \in \{1,..,N_C\}$}
        \State Obtain the input features vectors \(\bm{F}^{in}_{n_j}\) of  \(\bm{\theta^C}_j\) from \(\bm{F}^{in}_n\)  
        \Statex \hspace{9mm}and generate corresponding \(\bm{F}^{out}_{n_j}\) 
        \State Calculate the average distances \((d_j)\) between all \(\bm{f}^{out}_{n_j}\)  
        \Statex \hspace{9mm} in \(\bm{F}^{out}_{n_j}\) and their nearest neighbors in \(\bm{F}^{out}_b\)
    \EndFor

    \State Calculate the the average value \(\mu\) of  all \((d_j)\)
    \State \(j_1 = \argmax_{j} (d_j)\), \(j_2 = \argmin_{j} \mid d_j - \mu\mid\)
    \State U = \(\{\bm{\theta^C}_{j_1}, \bm{\theta^C}_{j_2}\}\), \(\bm{F}^{in}_u = \{\bm{F}^{in}_{n_{j_1}}, \textbf{F}^{in}_{n_{j_2}}\}\) 
    \State collecting query performance data according to U in Data 
    \Statex \hspace{4mm} Collector, and generating new training data \(\mathcal{T}_\text{u}\) 
    \State \(\mathcal{T} = \mathcal{T} \cup \mathcal{T}_\text{u}\), retrain the QPP model \(\phi\) using \(\mathcal{T}\) 
    \State \(\bm{\theta^C} = \bm{\theta^C} \setminus \text{U}\), \(\bm{F}^{in}_b = \bm{F}^{in}_b \cup \bm{F}^{in}_u\), \(\bm{F}^{in}_n = \bm{F}^{in}_n \setminus \bm{F}^{in}_u\)

    \State flag = \( \textbf{Data Similarity Detection}(\bm{F}^{in}_b,\bm{F}^{in}_n, \phi\))

    \If{flag is \(\textbf{True}\)}
        \State \(\textbf{break}\)
    \EndIf
\EndFor
\State \(\{DF_b\} = \{DF_b\} \cup \{DF_n\}\)
\State return \(\{DF_b\}\) and \(\phi\)
\renewcommand{\algorithmicrequire}{\textbf{Output}}
\Require the new base datasets' features \(\{DF_b\}\) and the updated QPP model \(\phi\)
\end{algorithmic}
\end{algorithm}

\par
In our context, the \(\bm{F}^{in}_b\) of base datasets are the initial labeled data, while \(\bm{F}^{in}_n\) of the new dataset are the initial unlabeled data. Additionally, we consider collecting data in a unit of construction parameter configuration rather than parameter configuration in the model transfer process. This is because compared to parameter configurations, collecting the same amount of data requires fewer construction configurations, which can avoid constructing extensive PGs. Therefore, the objective is to select construction configurations that can effectively minimize the distance between \(\bm{F}^{in}_n\) and \(\bm{F}^{in}_b\). To this end, we propose a construction parameter configuration selection (CPCS) algorithm, as outlined in Algorithm \ref{alg:cpcs}. In each iteration, the CPCS first initializes an empty for selected construction configurations and generates the output feature vectors \(\bm{F}^{out}_b\) corresponding to \(\bm{F}^{in}_b\) (Lines 2--3). It then obtains the output feature vectors for each candidate construction configuration from \(\bm{F}^{in}_n\) and calculates the average NND (ANND) between corresponding output feature vectors and \(\bm{F}^{out}_b\) (Lines 4--7). The CPCS selects the construction configurations with the largest and average ANNDs (Lines 8--10). This selection rationale is that we find Choosing only configurations with the maximum ANND often leads to similar selections, limiting the ability to reduce the distance between unlabeled and labeled data. In contrast, adding a configuration with average ANND increases diversity and improves effectiveness. Subsequently, the CPCS uses the Data Collector to collect query performance data for the selected construction configurations and retrains the QPP model (Lines 11--13). Finally, it calls the Data Similarity Detector to assess the similarity between the current labeled data and remaining unlabeled data (Lines 14--18). If similar, this process is terminated; Otherwise, the next iteration begins. The process repeats until the detection result indicates similarity or the maximum number of iterations is reached. This design aims to avoid selecting too many construction configurations. Upon completion of this process, the pre-trained QPP model is effectively transferred to the new dataset (Lines 20--21).

\section{EVALUATION}
In this section, we evaluate the tuning performance of PGTuner on HNSW \cite{HNSW}. In Section \ref{sec:tuning}, We compare PGTuner with baselines on six real-world datasets and conduct a successive transfer tuning study across four datasets to demonstrate its transferability. In Section \ref{sec:tune dynamic}, we further evaluate PGTuner's tuning ability in two scenarios where datasets are continuously changing. In Section \ref{sec:analysis}, We conduct a comprehensive study on the influence of various components of PGTuner and the scale of pre-training data on its performance. In Section \ref{sec:adapt}, we evaluate PGTuner on another representative PG, NSG \cite{NSG},  to verify its applicability to other PGs.
\par 
\noindent \textbf{Experimental Environment.} All experimental data are collected on two servers. Server1 has a 14-core 2.6GHz CPU and 224GB RAM, and Server2 has a 24-core 5.8GHz CPU and 128GB RAM. Both servers are equipped with an NVIDIA 4090 GPU. 
\par
\noindent \textbf{Datasets.} In the experiments, we use seven real-world datasets:   Nytimes\footnote{Nytimes: \url{https://github.com/erikbern/ann-benchmarks/?tab=readme-ov-file}}, Glove\footnote{Glove: \url{http://downloads.zjulearning.org.cn/data/glove-100.tar.gz}},  Tiny5M\footnote{Tiny5M: \url{https://www.cse.cuhk.edu.hk/systems/hash/gqr/datasets.html}}, Paper\footnote{Paper: \url{https://drive.google.com/file/d/1t4b93_1Viuudzd5D3I6_9_9Guwm1vmTn/view}}, Crawl\footnote{Crawl: \url{http://downloads.zjulearning.org.cn/data/crawl.tar.gz}}, Msong\footnote{Msong: \url{https://drive.google.com/file/d/1UZ0T-nio8i2V8HetAx4-kt_FMK-GphHj/view}}, GIST\footnote{GIST: \url{http://corpus-texmex.irisa.fr/}}, Deep10M\footnote{Deep10M: \url{http://sites.skoltech.ru/compvision/noimi/}}, and SIFT1B\footnote{SIFT1B: \url{http://corpus-texmex.irisa.fr/}}. We respectively extract 1M subsets from Deep10M and from SIFT1B as base datasets. Additionally, a 1M subset from Tiny5M and a 50M subset from SIFT1B are respectively extracted as the new datasets for tuning. Finally, we partition five base datasets and six new datasets. The basic information of these datasets and respective servers used are provided in Table \ref{tab:dataset}.
\begin{table}[tbp] 
    \centering 
    \caption{Evaluated Datasets.} 
    \label{tab:dataset} 
    \setlength{\tabcolsep}{4pt} 
    \renewcommand{\arraystretch}{1.2}
    \resizebox{\linewidth}{!}{ 
    \begin{tabular}{c|c|c|c|c|c} 
        \toprule[1pt]
        \textbf{Dataset} & \textbf{Size of BD} & \textbf{Size of QD} & \textbf{Dimension} & \textbf{Category} & \textbf{Server} \\  
        \midrule
        \midrule
        \rowcolor{gray!15}
        Deep1M & 1000000 & 10000 & 96 & Base & Server2\\ 
        SIFT1M & 1000000 & 10000 & 128 & Base & Server2 \\
        \rowcolor{gray!15}
        Paper & 2029997 & 10000 & 200 & Base & Server1 \\ 
        Crawl & 1989995 & 10000 & 300 & Base & Server2 \\ 
        \rowcolor{gray!15}
        Msong & 992272 & 200 & 420 & Base & Sever1 \\ 
       Nytimes & 290000 & 1000 & 256 & New & Sever2 \\ 
        \rowcolor{gray!15}
        Glove & 1183514 & 10000 & 100 & New & Server2 \\ 
        Tiny1M & 1000000 & 1000 & 384 & New & Sever2 \\
        \rowcolor{gray!15}
        GIST & 1000000 & 1000 & 960 & New & Server1 \\ 
        Deep10M & 10000000 & 10000 & 96 & New & Server1 \\ 
        \rowcolor{gray!15}
        SIFT50M & 50000000 & 10000 & 128 & New & Server2 \\ 
        \bottomrule[1pt]
    \end{tabular}
    }
\end{table}

\par
\noindent \textbf{Baselines.} PGTuner is compared with the following methods:
\begin{itemize}[left=0pt, labelsep=0.5em]
\item \textbf{GridSearch} \cite{wang2021comprehensive} constructs PGs according to predefined parameter configurations to search the best one.

\item \textbf{RandomSearch} uses Latin Hypercube Sampling (LHS) \cite{loh1996latin} to randomly sample a small number of construction parameter configurations to obtain the optimal parameter configuration.

\item \textbf{VDTuner} \cite{yang2024vdtuner} employs the Gaussian process regression and Constrained Bayesian optimization to auto-tune PG. 

\item \textbf{GMM} \cite{oyamada2020towards} pre-trains a Random Forest based-meta-model. For a new dataset, GMM collects query performance data with fixed parameter configurations to update the meta-model and then uses it to predict the query performance of all candidate configurations to select the best one.
\end{itemize}
\par

\noindent \textbf{Settings.} We consider 10-nearest neighbor search with Euclidean distance. For HNSW \cite{HNSW}, we selected $ efC$,$ M$, and $efS$ as the tuning parameters, which are key parameters of HNSW and have been widely considered in related researches \cite{ wang2021comprehensive, yang2024vdtuner, NSG}. Furthermore, according to \cite{HNSW, wang2021comprehensive} and preliminary experimental results, we set the ranges of $efC$, $M$, and $efS$ as [20, 800], [4, 100], and [10, 5000], with 20, 13, and 94 distinct values, respectively. For PGTuner, the selection rounds $R$ in Algorithm \ref{alg:cpcs} is set to 7 and the according to the experimental results. Moreover, the maximum number of recommendation rounds in the PCR model is set to 250. For RandomSearch, 14 construction configurations are randomly sampled to be consistent with PGTuner. For VDTuner, the number of tuning rounds is set to 50. Additionally, we sample 5 construction configurations and collect corresponding query performance data to initialize its surrogate model. For GMM, it uses the same pre-training data as PGTuner and 14 construction configurations are also sampled to update its meta-model on new datasets.
\par
\noindent \textbf{Metrics.} We evaluate the tuning performance of PGTuner and the baselines in terms of tuning effect and tuning efficiency. The tuning effect is measured by the QPS of the recommended configuration, while the tuning efficiency refers to the time taken for tuning (i.e., tuning time). Furthermore, we calculate the QPS improvements of PGTuner over the baselines to quantify the superiority of PGTuner's tuning effect. The QPS improvement is calculated as follows: 
\[
\Delta QPS = (QPS_P-QPS_b) / QPS_b \times 100 \%
\]
Where $\Delta QPS$ represents the QPS improvement; $QPS_P$ and $QPS_b$ denote the $QPS$ of PGTuner and a specific baseline, respectively.

\subsection{Tuning Performance Evaluation}\label{sec:tuning}
In this section, we evaluate the tuning performance of PGTuner and baselines on six real-world datasets. For Glove, Tiny1M and GIST datasets, we set 9 target recalls: 0.85, 0.88, 0.9, 0.92, 0.94, 0.95, 0.96, 0.98, and 0.99. For Nytimes dataset, the 0.99 recall is not considered because it is unreachable. For Deep10M and SIFT50M datasets, due to the longer tuning time on these large-scale datasets, we focus on three primary target recalls: 0.9, 0.95, and 0.99.
\begin{figure}[t]
  \centering
  \includegraphics[width=\linewidth]{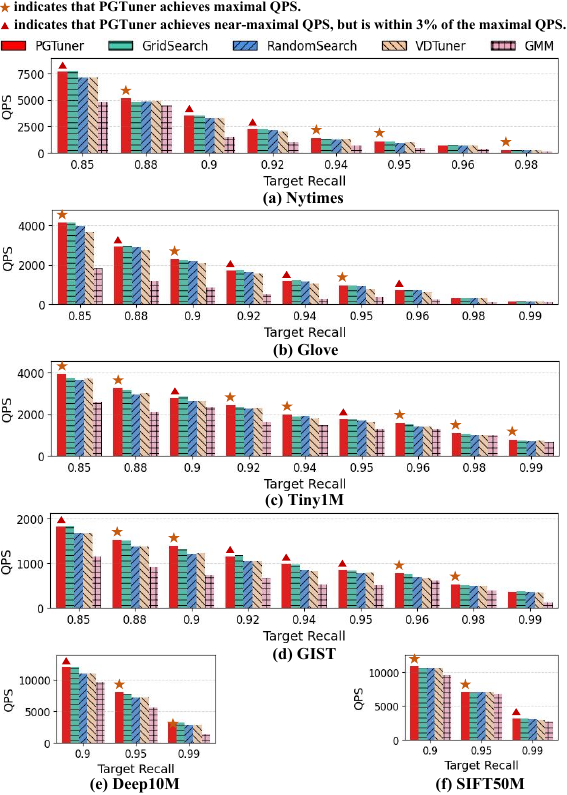}  
  \caption{ The QPS for all recommended parameter configurations at different target recalls.}
  \label{fig:test_main_qps}
\end{figure}

\subsubsection{Tuning Effect Comparison}\label{sec:tuning_effect}
We compare the tuning effect of PGTuner with all baselines. Figure \ref{fig:test_main_qps} reports the maximal QPS achieved by different methods on different datasets and target recalls. PGTuner achieves the highest QPS in 54\% of the dataset-recall combinations. This result demonstrates PGTuner has a stable and superior tuning ability across datasets of different scales and dimensions. Firstly, compared to GridSearch, the most competitive baseline, PGTuner achieves average QPS improvements of 1.58\%, -2.2\%, 2.26\%, 0.59\%, 2.76\%, and 1.28\% on Nytimes, Glove, Tiny1M, GIST, Deep10M, and SIFT50M datasets. It can be seen that PGTuner stably reaches a tuning effect similar to or better than that of GridSearch overall. This is because PGTuner learns a high-quality and generalizable tuning strategy through pre-training, enabling it to recommend better configurations in most cases. Secondly, PGTuner generally outperforms all other baselines. It achieves average QPS improvements of 2.16\%-12.91\% compared to RandomSearch. It can be observed that RandomSearch produces relatively worse tuning effects on datasets like Nytimes, Tiny1M, GIST, and Deep10M. The reason for its varying performance across datasets is that it samples configurations independently of dataset, preventing it from consistently finding high-performance configurations for each dataset. Furthermore, PGTuner also achieves average QPS improvements of 3.9\%-13.73\% over VDTuner. The suboptimal tuning effect of VDTuner arises from its inability to accurately capture complex dependencies between configurations, query performance, and tuning objectives, leading to missing promising configurations and unstable results. Similarly, GMM performs poorly on datasets other than SIFT50M as it fails to effectively generalize the meta-model and capture the complex dependencies, resulting in large prediction errors that hinder accurate configuration recommendations.
\begin{figure}[t]
  \centering
  \includegraphics[width=\linewidth]{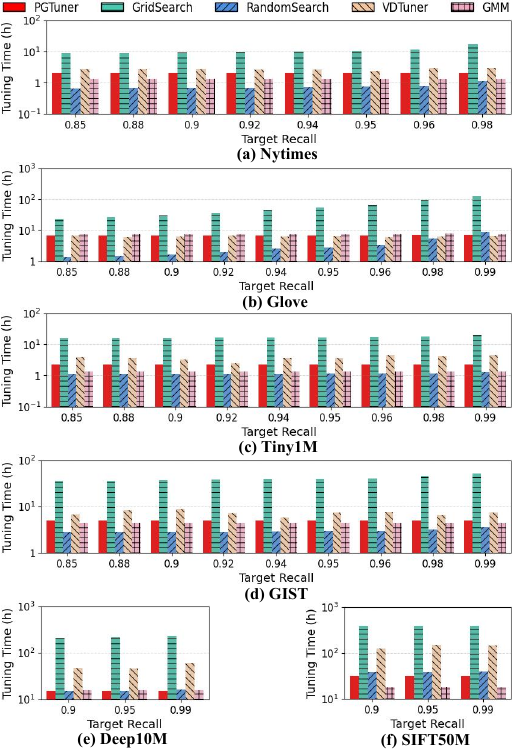}  
  \caption{ The tuning time at different target recalls.}
  \label{fig:test_main_time}
\end{figure}

\subsubsection{Tuning Efficiency Comparison}
In this section, we compare the tuning efficiency of PGTuner with baselines. Figure \ref{fig:test_main_time} displays the tuning time of each method. Firstly, it can be observed that PGTuner's tuning time is similar across different target recalls on the same dataset because it can parallel tune for multiple target recalls. Secondly, PGTuner is on average 5.11$\times$, 7.91$\times$, 7.4$\times$, 8.16$\times$, 14.69$\times$ and 12.34$\times$ faster than GridSearch on Nytimes, Glove, Tiny1M, GIST, Deep10M and SIFT50M respectively. This result shows that PGTuner can significantly reduce the tuning time while stably achieving the similar tuning effect as GridSearch, and its advantage generally increases as the size and dimension of the dataset expand. In contrast, PGTuner achieves average speedups of 0.36-1.22 over RandomSearch on these datasets. RandomSearch tunes faster on Nytimes, Glove, Tiny1M, and GIST due to the short data collection time resulting from their small sizes or low dimensions, especially at low target recalls. Although PGTuner is slower on these datasets, its tuning time is still short while attaining a higher-level tuning effect. Compared with VDTuner, PGTuner improves the tuning efficiency by an average of 0.92$\times$-4.5$\times$. VDTuner is slightly faster on Glove also because of the low dimension and small size of Glove, leading to short construction time of HNSW. The tuning time between PGTuner and GMM is mostly minimally different, as they both collect query performance data of 14 construction configurations during tuning, which dominates the tuning time. 
\par
Besides efficiently tuning for a fixed target recall, PGTuner boosts tuning efficiency more significantly as user recall preference rises. For example, when increasing the target recall from 0.95 to 0.99, PGTuner only needs to fine-tune the PCR model to quickly recommend a new optimal configuration. In contrast, GridSearch and RandomSearch must re-execute search on all constructed PGs (assuming constructed PGs are stored) to find new optimal configuration for 0.99 recall. At this time, PGTuner is 16.89$\times$-125.17$\times$ faster than GridSearch and 1.44$\times$-10.13$\times$ faster than RandomSearch on Glove, GIST, Deep10M and SIFT50M datasets. These results further demonstrate PGTuner’s superior tuning efficiency and flexibility.

\begin{figure}[t]
  \centering
  \includegraphics[width=\linewidth]{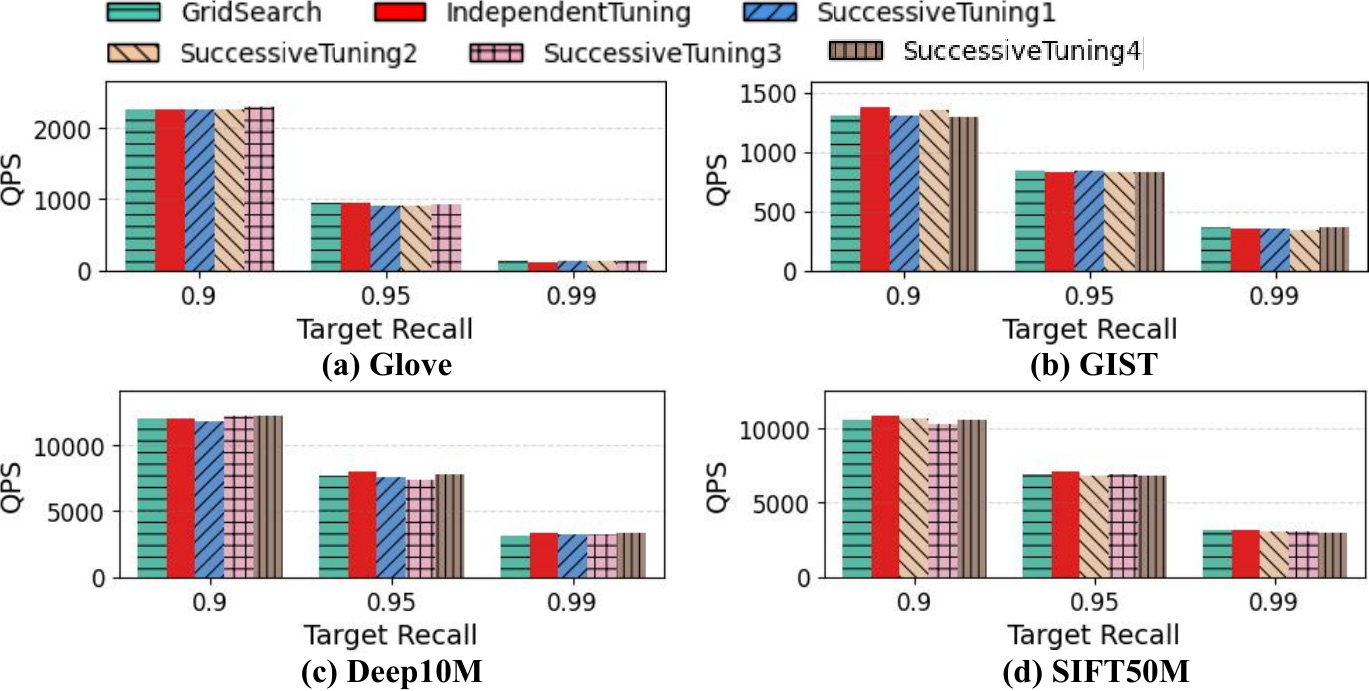}  
  \caption{The comparison of QPS for independent and successive tuning at different target recalls.}
  \label{fig:successive_QPS}
\end{figure}

\subsubsection{Successive Transfer Tuning Study}
In Section 5.1.1, PGTuner tunes independently on four datasets without leveraging the previous tuning knowledge. However, in practice, as a vector database may continuously store new datasets, tuning on each new dataset is often performed based on previous tuning. Therefore, we use PGTuner to perform tuning across four datasets in succession to evaluate its performance for transfer tuning, considering target recalls of 0.9, 0.95, and 0.99. We set up four tuning sequences: (1) SIFT50M, Deep10M, GIST, Glove; (2) Deep10M, GIST, Glove, SIFT50M; (3) GIST, Glove, SIFT50M, Deep10M; and (4) Glove, SIFT50M, Deep10M, GIST. For each dataset, we compare the tuning effect of independent tuning with that of successive tuning in three sequences, where the dataset is not at the first position in the sequence.
\par
Figure \ref{fig:successive_QPS} shows the tuning effect of independent tuning and successive tuning of different tuning sequences. In the legend, IndependentTuning represents independent tuning, while SuccessiveTuning1 to SuccessiveTuning4 correspond to the four sequences. First, it can be seen that the tuning effect of successive tuning across different tuning sequences reaches or exceeds that of GridSearch. Secondly, successive tuning generally aligns with independent tuning, except for an improvement up to 11.51\% at the target recall of 0.99 on Glove. This is because the knowledge gained from previous tuning helps the QPP model to be more effectively transferred to Glove, which differs significantly from the base datasets, enabling the PCR model to accurately recommend high-performance configurations. These results prove the powerful transferability of PGTuner, which can help it maintain stable and superior tuning effect in successive transfer tuning.

\subsubsection{Overhead Analysis}
In this section, we provide the time overhead of PGTuer during pre-training and online tuning, which are presented in Table \ref{tab:pre_overhead} and Table \ref{tab:online_overhead}. The tuning process of PGTuner mainly includes five stages: feature extraction, data collection, QPP model training, PCR model training and post-processing. Since the post-processing time is very short (typically less than one minute) and associated with the specific target recall. Therefore, it is not shown in Table \ref{tab:online_overhead} but is included in the tuning time calculation.

\begin{table}[tbp] 
    \centering 
    \caption{The time overhead (s) of PGTuner during the pre-training phase.} 
    \label{tab:pre_overhead} 
    \setlength{\tabcolsep}{9pt}
    \renewcommand{\arraystretch}{1.1}
    \resizebox{\linewidth}{!}{ 
    \begin{tabular}{c|>{\centering\arraybackslash}m{1.5cm}|>{\centering\arraybackslash}m{1.5cm}|>{\centering\arraybackslash}m{1.5cm}|>{\centering\arraybackslash}m{1.5cm}}
        \toprule[1pt]
        \textbf{Dataset} & \textbf{Feature} \newline \textbf{Extraction} & \textbf{Data} \newline \textbf{Collection} & \textbf{QPP} \newline \textbf{Training} & \textbf{PCR} \newline \textbf{Training} \\
        \midrule
        \midrule
        \cellcolor{gray!15} Deep1M &\cellcolor{gray!15} 38.6 & \cellcolor{gray!15} 61136.1 & \multirow{5}{*}{561.6} & \multirow{5}{*}{27643.3} \\ 
        SIFT1M & 49.7 & 51475.2 &  &  \\ 
        \cellcolor{gray!15} Paper & \cellcolor{gray!15} 213.7 & \cellcolor{gray!15} 129367.9 &  &  \\ 
        Crawl & 242.8 & 428819.5 &  &  \\ 
        \cellcolor{gray!15} Msong & \cellcolor{gray!15} 110.3 & \cellcolor{gray!15} 49668.4 &  &  \\ 
        \midrule
        \multicolumn{5}{r}{\textbf{Total: 749327.1}} \\ 
        \bottomrule[1pt]
    \end{tabular}
    }
\end{table}

\begin{table}[tbp] 
    \centering 
    \caption{The time overhead (s) of PGTuer on each dataset during the online tuning phase.} 
    \label{tab:online_overhead} 
    \setlength{\tabcolsep}{4pt}
    \renewcommand{\arraystretch}{1.1}
    \resizebox{\linewidth}{!}{ 
    \begin{tabular}{c|>{\centering\arraybackslash}m{1.5cm}|>{\centering\arraybackslash}m{1.5cm}|>{\centering\arraybackslash}m{1.5cm}|>{\centering\arraybackslash}m{1.5cm}|c}
    \toprule[1pt]
    \textbf{Dataset} & \textbf{Feature} \newline \textbf{Extraction} & \textbf{Data Collection} & \textbf{QPP Training} & \textbf{PCR Training} & \textbf{Total} \\
    \midrule
    \midrule
    \rowcolor{gray!15}
    Ntimes & 9.9 & 3303.8 & 3308.9 & 865 & 7487.6 \\ 
     Glove & 44.8 & 20268.5 & 3459.1 & 833 & 24605.4 \\ 
    \rowcolor{gray!15}
    Tiny1M & 168.3 & 3792.9 & 3314.6 & 864 & 8139.8 \\ 
    GIST & 419.3 & 8896 & 3359.4 & 840 & 13514.7 \\ 
    \rowcolor{gray!15}
    Deep10M & 942.8 & 32970.2 & 3298.9 & 853 & 38064.9 \\ 
    SIFT50M & 10116.7 & 99718 & 3339.9 & 676.9 & 113851.5 \\
    \bottomrule[1pt]
\end{tabular}
}
\end{table}

\par
Since the data are collected on different servers, we cannot directly calculate the total time overhead. According to data statistics, the computation speed of Server2 is approximately 1.47x faster than that of Server1. Therefore,  we approximate the data collection time on Server1 to the corresponding time on Server2 in Table \ref{tab:pre_overhead} and Table \ref{tab:online_overhead}. It can be seen that the time overhead of pre-training is 749327.1 seconds (about 208.1 hours). Although this overhead is relatively long, it can be quickly amortized in the tuning on new datasets, especially on tens-of-millions-scale datasets. For example, the total overhead of pretraining and tuning on Glove, GIST and Deep10M is 227.8 hours, which is lower than the average tuning time of GridSearch on these three datasets (240.2 hours). This result further strongly demonstrates that PGTuner is highly efficient. 

\subsection{Tuning Performance Comparison in Dynamic Scenarios}\label{sec:tune dynamic}
In real-world scenarios, a dataset may change over time, such as the increasing data size of the BD or the variable QD, making the current optimal configuration unable to achieve the target recall. In such cases, it is necessary to re-tune the PG. In this section, we simulate the two dynamic scenarios described above to further evaluate the tuning performance of PGTuner at the target recall of 0.95. We assume that the dataset changes at a low frequency so that all methods can complete the current tuning before any new changes occur in the dataset. Achieving stable real-time tuning in a highly dynamic environment is beyond the scope of this work and will be the focus of future research.

\subsubsection{Comparison under the Continuous Increase in BD Size}
\begin{figure}[tbp]
  \centering
  \includegraphics[width=\linewidth]{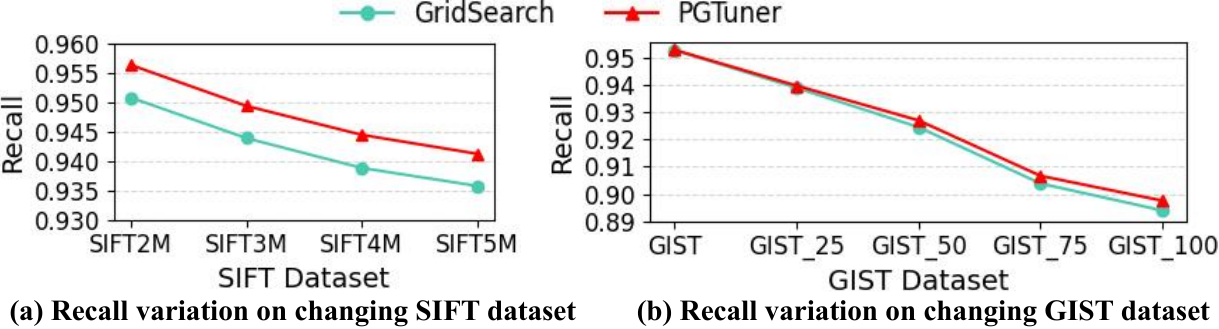}  
  \caption{The variations in query recall of the initially tuned configurations under increasing BD and changing QD.}
  \label{fig:ds_qd_recall}
\end{figure}

In this section, we extract subsets with data sizes of 5M, 4M, 3M, and 2M from SIFT50M to simulate the continuous increase in data size of BD. The corresponding datasets are named SIFT5M, SIFT4M, SIFT3M, and SIFT2M. As shown in Figure \ref{fig:ds_qd_recall} (a), the optimal configurations recommended by GridSearch and PGTuner for SIFT2M yield recalls lower than 0.95, with a continuous decrease. Therefore, we perform tuning sequentially from SIFT2M to SIFT5M. 
\par
\begin{figure}[tbp]
  \centering
  \includegraphics[width=\linewidth]{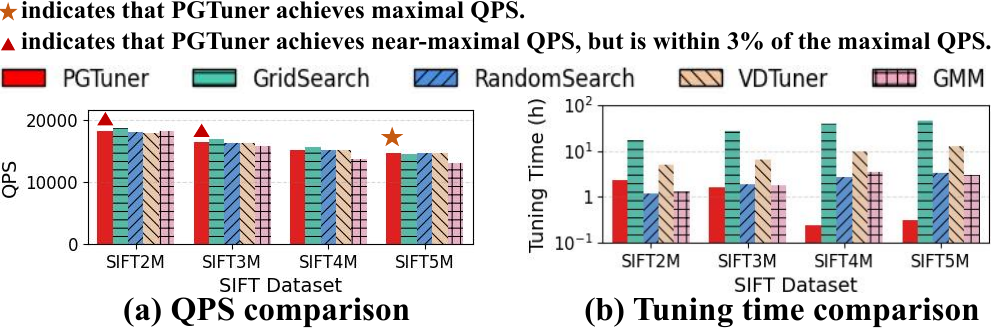}  
  \caption{The QPS and Tuning time on SIFT datasets with increasing data size of BD.}
  \label{fig:ds_QPS_time}
\end{figure}

Figure \ref{fig:ds_QPS_time} (a) shows the maximal QPS achieved by each method on the continuously growing SIFT datasets. It can be observed that PGTuner reaches the near-optimal or optimal tuning effect on all SIFT datasets except SIFT4M. Except for being slightly inferior to GridSearch, the tuning effect of PGTuner is better than that of other baselines. This result shows that PGTuner can maintain excellent tuning capability when the size of BD continuously increases. 
\par
Figure \ref{fig:ds_QPS_time} (b) displays the tuning time of each method on all datasets. It can be seen that PGTuner achieves the fastest tuning on SIFT3M-SIFT5M, especially on SIFT4M and SIFT5M. On SIFT3M-SIFT5M, PGTuner is 16.77$\times$-169.73$\times$ faster than GridSearch, 1.18$\times$-11.55$\times$ faster than RandomSearch, 4$\times$-42.22$\times$ faster than VDTuner, and 1.15$\times$-14.64 times faster than GMM, respectively. The significant acceleration of PGTuner can primarily be attributed to the Data Similarity Detector (DSD). When PGTuner performs tuning on the SIFT3M, it first iteratively collects query performance data to update the current QPP model. However, the DSD detects the similarity between SIFT3M and the current base datasets after three iterations.  As a result, the QPP model transfer is early terminated, which helps effectively reduce the amount of data collected during model transfer, thereby greatly accelerating the overall tuning process. Furthermore, when starting tuning on SIFT4M and SIFT5M, the DSD identifies the similarities during initial detection. Consequently, the QPP model does not need to be updated, and the time-consuming model transfer process is bypassed. PGTuner can then quickly recommend the optimal configuration by fine-tuning the PCR model, achieving an efficient tuning. This outcome validates the effectiveness of our designed DSD and also indicates that PGTuner could achieve higher tuning efficiency in the dynamic scenario where the BD size continuously increases.

\begin{figure}[tbp]
  \centering
  \includegraphics[width=\linewidth]{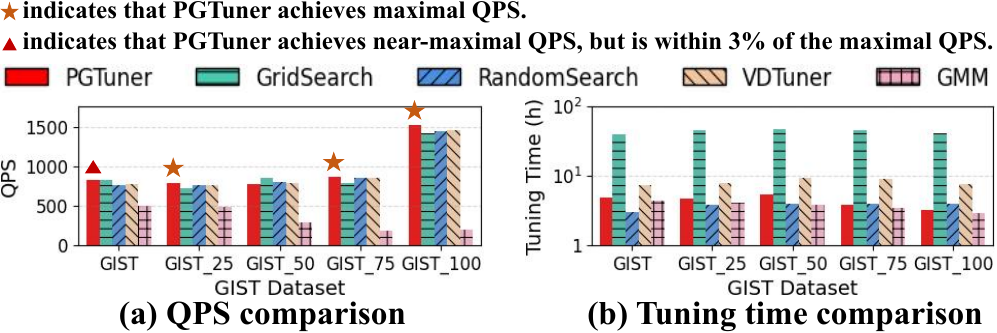}  
  \caption{The QPS and Tuning time on GIST datasets with the changes of QD.}
  \label{fig:qd_qps_time}
\end{figure}
\subsubsection{Comparison under the Continuous Changes in QD}
In this section, We simulate the continuous variation of QD in the GIST dataset by adding Gaussian noise at levels of 25\%, 50\%, 75\%, and 100\%, and the corresponding datasets are named GIST\_25, GIST\_50, GIST\_75, and GIST\_100 respectively. As shown in Figure \ref{fig:ds_qd_recall} (b), The optimal configuration tuned on GIST results in recalls below 0.95 after GIST’s QD varies, with a rapid decline as the degree of QD change increases. As a result, after tuning on GIST, we sequentially perform tuning from GIST\_25 to GIST\_100.
\par
The QPS comparison is shown in Figure \ref{fig:qd_qps_time} (a). As can be seen, PGTuner shows the best tuning effect on all datasets except on GIST\_50, where it performs relatively worse. This result demonstrates that PGTuner also can stably achieve  advanced tuning effect in the dynamic scenario where the QD changes continuously. Figure \ref{fig:qd_qps_time} (b) shows the comparison of tuning times. Although the Data Similarity Detector fails to effectively shorten the tuning time of PGTuner due to the large variation in QD, PGTuner still achieves near-maximum tuning efficiency. 

\subsection{Analysis of PGTuner}\label{sec:analysis}
In this section, we first evaluate the effectiveness of the PCR model and Constrction Parameter Configurations Selection (CPCS) algorithm of PGTuner through ablation study. Then, we investigate the impact of two key parameters in the CPCS algorithm on the performance of the updated QPP model. After that, we study the impact of the parameter $k$ in the Feature Extractor on PGTuner's tuning effect. Finally, we conduct an exploration of the relationship between different sizes of pretraining data and PGTuner's tuning performance to determine the best pretraining data size.

\begin{figure}[tbp]
  \centering
  \includegraphics[width=\linewidth]{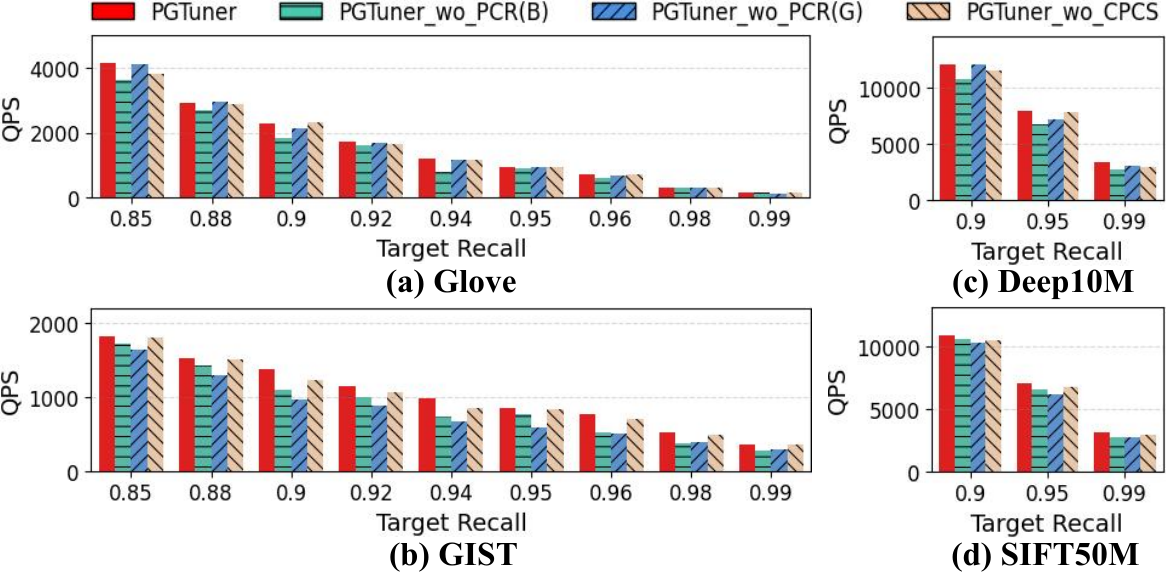}  
  \caption{The QPS of PGTuner and its variants.}
  \label{fig:ablation_QPS}
\end{figure}

\subsubsection{Ablation Study}
To evaluate the effectiveness of the two key components, we replace one component with a simple method while keeping the other fixed. For the PCR model, we replace it with the Bayesian optimization (BO) and the recommendation method from GMM respectively, and name the corresponding variants as PGTuner\_wo\_PCR(B) and PGTuner\_wo\_PCR(G). The tuning time of PGTuner\_wo\_PCR(B) on each dataset is set the same as PGTuner's. For the CPCS, we use random sampling as a replacement, denoted as PGTuner\_wo\_CPCS. 
\par
Figure \ref{fig:ablation_QPS} presents the QPS of PGTuner and its three variants. PGTuner outperforms all variants in most cases and achieves average improvements of up to 24.24\%, 33.73\% and 10.58\%, compared to PGTuner\_wo\_PCR(B), PGTuner\_wo\_PCR(G) and PGTuner\_wo\_CPCS respectively. These results validate the effectiveness of the PCR model and CPCS algorithm, as removing either will significantly harm PGTuner’s tuning effect. Additionally, PGTuner\_wo\_CPCS generally shows higher QPS than PGTuner\_wo\_PCR(B) and PGTuner\_wo\_PCR(G), indicating that the PCR model makes a greater contribution to achieving the superior tuning ability of PGTuner. For PGTuner\_wo\_PCR(B), the BO may be misled to deviate from the global optimum due to the QPP model's prediction errors, making it prone to yield suboptimal and unstable tuning results. In contrast, the DRL-based PCR model learns a generalizable and high-quality tuning strategy through pre-training, enabling PGTuner to achieve a more stable and excellent tuning. The comparison of tuning times is not shown because the differences between them are minimal.

\begin{figure}[tbp]
  \centering
 \includegraphics[width=0.98\linewidth]{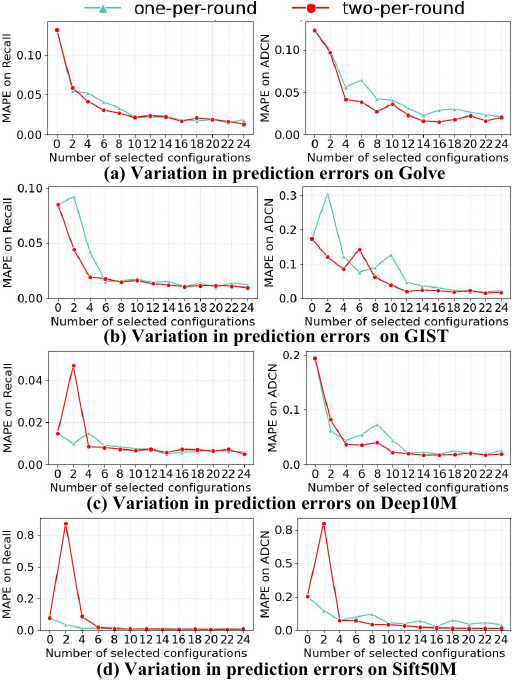}  
  \caption{The variations of MAPEs on recall and ADCN with respect to the number of selected configurations under two selection strategies.}
  \label{fig:parameter_test_errors}
\end{figure}

\subsubsection{CPCS Algorithm Study}
In this section, we explore the effects of both the number of selection rounds and the number of configurations selected per round in the CPCS algorithm on the QPP model's prediction performance. The construction configurations selected per round can be set to 1 or 2, corresponding to two selection strategies: one-per-round and two-per-round. The one-per-round strategy selects the configuration with the maximum nearest neighbor (NN) distance to the labeled data per round, while the two-per-round strategy selects two configurations with the largest and average NN distances per round. We set the maximum number of selectable configurations to 24.
\par
Figure \ref{fig:parameter_test_errors} shows the variations in MAPEs for recall and ADCN as the number of selected configurations increases under two selection strategies. The x-axis value of 0 represents the initial prediction errors of the pre-trained QPP model. It can be seen that both strategies quickly reduce prediction errors to a low level, demonstrating the effectiveness of the CPCS algorithm in model transfer. Additionally, the two-per-round strategy generally leads to faster and larger reductions in MAPEs on both recall and ADCN compared to the one-per-round strategy. Moreover, with the same number of selected configurations, the one-per-round strategy requires twice as many rounds, resulting in double the model retraining time. Therefore, the two-per-round strategy is more effective and efficient. 
\par
According to the MAPEs variations under the two-per-round strategy, it can be seen that when the number of selected configurations reaches 14 (i.e., the number of selection rounds is 7), the MAPEs essentially stabilize at or near their minimum values, with minimal change in further rounds. Thus, the number of selection rounds for the two-per-round strategy is set to 7 in consideration of minimizing the tuning time of PGTuner.

 \begin{figure}[tbp]
  \centering
  \includegraphics[width=\linewidth]{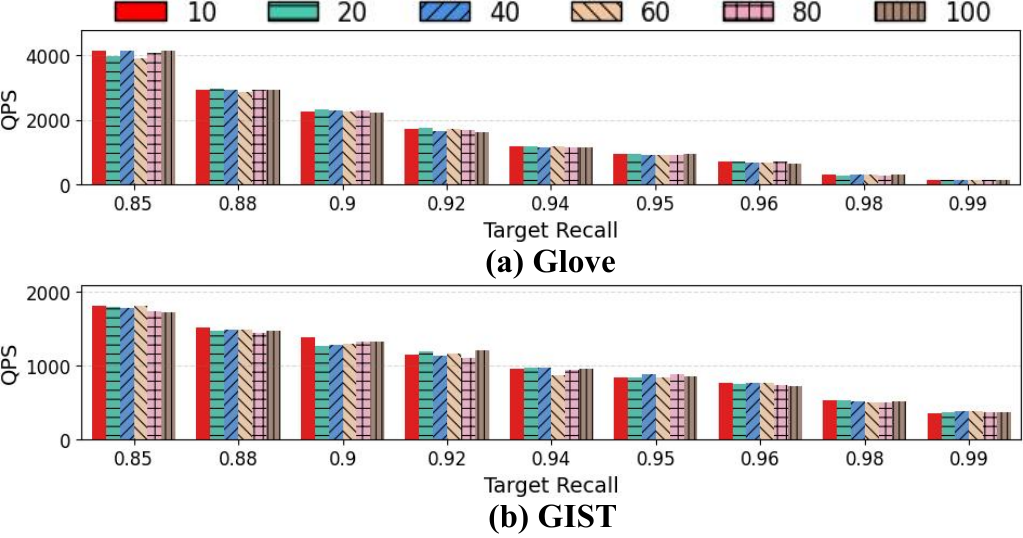}  
  \caption{The comparison of tuning effects of PGTuner under different values of $k$.}
  \label{fig:k_value_qps}
\end{figure}

\subsubsection{Feature Exactor Study}
In the Feature Extractor, parameter $k$ is used to calculate $DS$ and $DR$ features, which may affect the prediction performance of the QPP model and in turn influences PGTuner's tuning effect. We set the values of $k$ as 10, 20, 40, 60, 80, and 100, respectively. We then calculate the $DS$ and $DR$ features with different $k$ values and train corresponding QPP and PCR models. Then tuning tests are conducted on Glove and GIST datasets. The PGTuner corresponding to a specific $k$ is denoted as PGTuner\_$k$.
\par
Figures \ref{fig:k_value_qps} displays the tuning effects of different PGTuners. On Glove dataset, PGTuner\_10 outperforms PGTuner\_20-PGTuner\_100 with average QPS improvements of 1.51\%, 1.24\%, 0.81\%, 2.46\%, and 2.89\% respectively, and the improvements on GIST are 0.44\%, 0.12\%, 1.61\%, 2.29\%, and 1.27\% respectively. These results suggest that PGTuner\_10 possesses the best tuning effect. Meanwhile, the tuning effects of all PGTuners vary little overall, indicating that the tuning effect of PGTuner can remain stable across different $k$ values. Additionally, a larger $k$ typically results in higher time and memory costs to calculate $DS$ and $DR$. Therefore, setting $k$ to 10 is the best choice for PGTuner.

\begin{figure}[tbp]
  \centering
  \includegraphics[width=\linewidth]{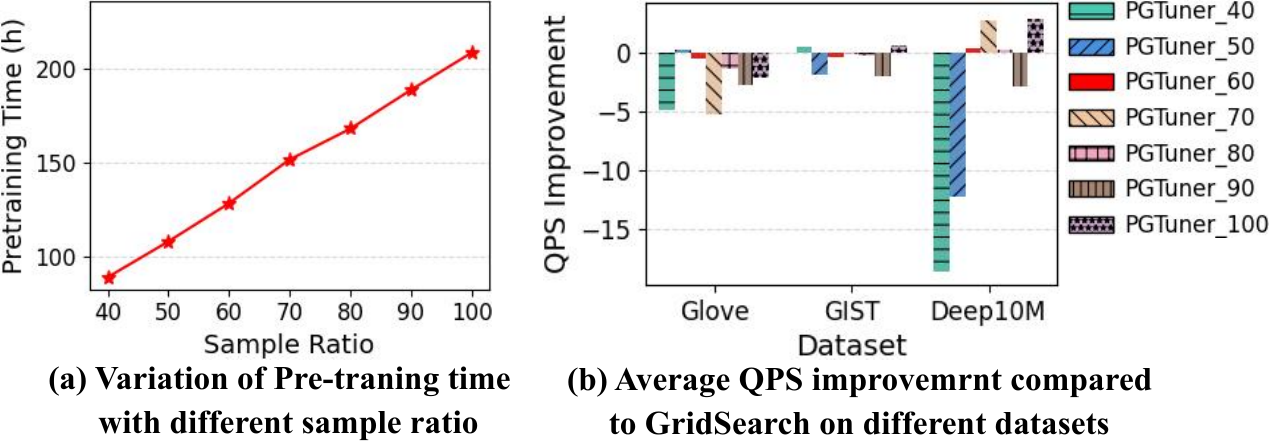}  
  \caption{The comparison of pre-training time and tuning effects of PGTuner under pre-training data of different size.}
  \label{fig:data_size_qps}
\end{figure}

\subsubsection{Size of Pre-training Data Study}
The size of pre-training data affects PGTuner's tuning effect. More pre-training data usually leads to higher-quality pre-trained QPP and PCR models, enhancing PGTuner's tuning ability. However, this also incurs higher pre-training time. Therefore, we study the impact of the pre-training data size on PGTuner's tuning effect. The goal is to find the best data size that can maintain the superior tuning effect of PGTuner while minimizing its pre-training time. We employ LHS to sample 40\%-90\% of configurations at 10\% intervals from all construction configurations and retrieve corresponding pre-training data. Subsequently, we execute PGTuner based on each sampled pre-training data and conduct tuning tests on Glove, GIST, and Deep10M datasets.
\par
Figure \ref{fig:data_size_qps} (a) shows the pre-training times corresponding to the sampled pre-training data at different ratios, which are 89.4, 108.3, 128.5, 151.6, 167.9, 188.4, and 208.1 hours respectively, showing a rapid linear increase with data size growth. Figure \ref{fig:data_size_qps} (b) illustrates the average QPS improvements of PGTuners over GridSearch based on different ratios. The PGTuner corresponding to the ratio of r\% is denoted as PGTuner\_r. The PGTuner\_100 represents the PGTuner executed on the original pre-training data. Firstly, it can be seen that PGTuner\_40 and PGTuner\_50 exhibit worse tuning effect, especially on Deep10M dataset. In contrast, PGTuner\_100 achieves the best tuning effect, which aligns with our analytical expectations. Moreover, both PGTuner\_60 and PGTuner\_80 demonstrate stable tuning effect on par with GridSearch. Therefore, according to these results, using pre-training data sampled at a 60\% ratio can sustain PGTuner's superior tuning ability while significantly reducing the pre-training time by 38.25\%. 

\subsection{Adaptability}\label{sec:adapt}
In this section, we evaluate PGTuner on NSG \cite{NSG}. According to \cite{NSG, wang2021comprehensive} and preliminary experimental results, we consider six key parameters of NSG: $K$, $L$, $L\_nsg$, $R\_nsg$, $C$, and $L\_s$. The ranges of these parameters are set to [100, 400], [100, 400], [150, 350], [5, 90], [300, 600], and [10, 1500], containing 4, 7, 5, 11, 4, and 52 distinct values, respectively. Specifically, $K$, $L$, $L\_nsg$, $R\_nsg$ and $C$ are construction parameters, while $L\_s$ is the search parameter.  For PGTuner, the parameter $R$ is set to 6, thus selecting up to 192 construction configurations. Similarly, 192 construction configurations are also randomly sampled for both RandomSearch and GMM. For VDTuner, the number of tuning iterations and initialized constructions configuration is set to 200 and 30 respectively. We use Deep1M and Paper as base datasets and GIST for tuning. Due to the long construction time of NSG, we extract subsets containing one-tenth of the data from these datasets, named Deep100K, Paper200K, and GIST100K respectively. The optimal configuration tuned on GIST100K is applied to GIST to measure the corresponding QPS. Three primary target recalls are considered: 0.9, 0.95, and 0.99.

 \begin{figure}[tbp]
  \centering
  \includegraphics[width=\linewidth]{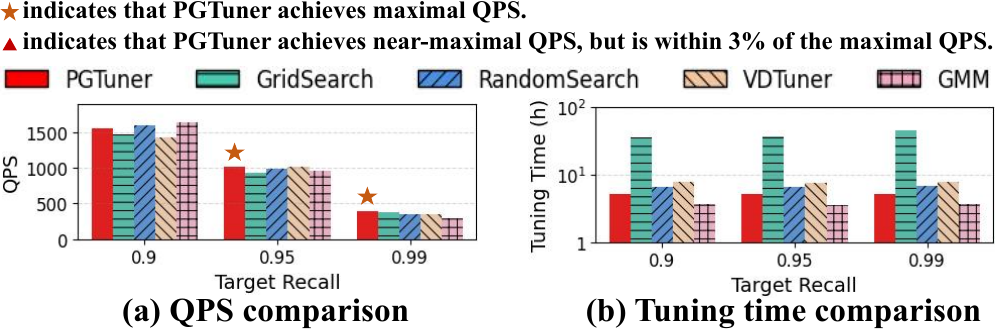}  
  \caption{The QPS and Tuning time on GIST dataset.}
  \label{fig:NSG_QPS_time}
\end{figure}

\par
Figure \ref{fig:NSG_QPS_time} (a) shows the tuning effect of different methods. PGTuner achieves improvements of up to 8.83\%, 15.99\%, 14.99\%, and 32.56\% over GridSearch, RandomSearch, VDTuner, and GMM, respectively. Figure \ref{fig:NSG_QPS_time} (b) shows that PGTuner also achieves higher tuning efficiency though it is 1.4$\times$ slower than GMM. However, this difference is acceptable given the significant improvement in the tuning effect. These results demonstrate that PGTuner is highly applicable for configuration tuning on other PGs.

\section{RELATED WORK}
\subsection{Automatic Configuration Tuning for PGs}
The current methods for automatic PG configuration tuning can be divided into search-based methods \cite{wang2021comprehensive, zhou4734062auto} and learning-based methods \cite{oyamada2020towards, yang2024vdtuner}.
\par
\noindent \textbf{Search-based.} GridSearch \cite{wang2021comprehensive} is widely used to search the optimal configurations by constructing numerous PGs, which is extremely costly. IGS-HNSW \cite{zhou4734062auto} reduces the number of constructed PG by leveraging the relationship between the average out-degree of HNSW and its query performance. However, IGS-HNSW is only applicable to tuning construction configuration for HNSW.
\par
\noindent \textbf{Learning-based}. VDTuner \cite{yang2024vdtuner} leverages the Constrained Bayesian optimization to iteratively recommend promising PG configurations. However, it is not enough efficient due to the requirement for constructing PGs during the tuning process. GMM \cite{oyamada2020towards} pre-trains a random forest-based meta-model to predict the query performance of candidate configurations on new datasets and recommends the best configurations. Although learning-based methods generally achieve faster tuning, they cannot transfer to new datasets effectively and fail to accurately capture the complex dependencies between PG configurations, query performance, and tuning objectives, making it difficult for them to reliably achieve outstanding performance in both tuning effect and efficiency simultaneously.

\subsection{Out-of-Distribution Detection} 
The Out-of-Distribution (OOD) detection aims to detect whether the test data falls within the distribution of the training data. There are mainly three categories of methods for OOD detection \cite{yang2024generalized}. Classification-based methods use the maximum softmax probability to detect classification models \cite{hendrycks2016baseline, liu2020energy, lin2021mood}. However, they are not suitable for the QPP model which belongs to the regression model. Density-based methods model the data distribution and classify samples with low density as OOD \cite{kobyzev2020normalizing, ren2019likelihood, xiao2020likelihood}. However, these methods require distributional assumptions and have high computational costs. Distance-based methods compute distances between test and training samples \cite{lee2018simple, ren2021simple, sun2022out}. \cite{lee2018simple, ren2021simple} use Mahalanobis distance for detection, which still requires assumptions about the feature space distribution. In contrast, \cite{sun2022out} calculates $K$ nearest neighbor (NN) distance between the test and training samples, which is efficient and matches our work. Thus, we choose it for our data similarity detection.

\subsection{Deep Active Learning}
Deep active learning aims to achieve strong performance with fewer training samples. It iteratively selects unlabeled samples for labeling based on a core query strategy. Current query strategies are mainly categorized into uncertainty-based, influence-based, and representativeness-based methods \cite{li2024survey}. Uncertainty-based methods select samples with the highest uncertainty \cite{xie2022active, kothawade2022talisman, elenter2022lagrangian}. However, these methods are often task-specific, limiting their generalizability. Influence-based methods select samples with the greatest impact on the model's performance \cite{gudovskiy2020deep, zhao2021efficient, mohamadi2022making}, which are computationally expensive. Representativeness-based methods aim to select samples that effectively cover the feature space \cite{yang2022actively, sener2017active}. \cite{yang2022actively} clusters samples and selects the cluster centers, which is very costly for large-scale and high-dimensional vectors. In contrast, \cite{sener2017active} selects samples with the largest NN distance to labeled samples, offering higher generality and computational efficiency, and it also aligns well with our work. Therefore, we improve it to enable model transfer.

\section{CONCLUSION}
In this paper, we propose PGTuner, an automatic and transferable PG configuration tuning framework. PGTuner can effectively and efficiently recommend high-performance configurations that satisfy users' recall requirements on any given dataset. Extensive experiments demonstrate that PGTuner significantly outperforms baselines in tuning performance. Moreover, it also achieves better tuning performance in dynamic scenarios.



\bibliographystyle{ACM-Reference-Format}
\balance
\bibliography{references}

\end{document}